\documentclass[twocolumn,showpacs,preprintnumbers,amsmath,amssymb]{revtex4}

\usepackage{graphicx}
\usepackage{dcolumn}
\usepackage{bm}
\usepackage{epsfig}

\begin{document}

\title{Enhanced thermal stability and spin-lattice relaxation rate of N@C$_{60}$ inside carbon
nanotubes}
\author{S. T\'oth $^{1}$, D. Quintavalle$^{1}$, B. N\'{a}fr\'{a}di$^{2}$, L. Korecz$^{3}$, L. Forr\'{o}$^{2}$, and F. Simon$^{1,\ast}$}

\affiliation{$^{1}$ Budapest University of Technology and Economics,
Institute of Physics and Condensed Matter Research Group of the
Hungarian Academy of Sciences, H-1521, Budapest P.O.Box 91, Hungary}
\affiliation{$^{2}$ Institute of Physics of Complex Matter, FBS
Swiss Federal Institute of Technology (EPFL), CH-1015 Lausanne,
Switzerland} \affiliation{$^{3}$ Chemical Research Center, Institute
of Chemistry, P.O. Box 17, H-1525 Budapest, Hungary}

\date{\today}


\begin{abstract}
We studied the temperature stability of the endohedral fullerene
molecule, N@C$_{60}$, inside single-wall carbon nanotubes using
electron spin resonance spectroscopy. We found that the nitrogen
escapes at higher temperatures in the encapsulated material as
compared to its pristine, crystalline form. The temperature
dependent spin-lattice relaxation time, $T_1$, of the encapsulated
molecule is significantly shorter than that of the crystalline
material, which is explained by the interaction of the nitrogen spin
with the conduction electrons of the nanotubes.
\end{abstract}
\maketitle

\section{Introduction}

Fullerenes encapsulated inside single-wall carbon nanotubes
(SWCNTs)\cite{IijimaNat1993,BethuneNat1993} is an interesting
molecular nanostructure as it combines two fundamental forms of
carbon. The existence of this so-called peapod structure
\cite{SmithNat} inspired a number of fundamental and application
oriented studies. The growth of peapods was studied with molecular
dynamics simulations \cite{TomanekPRL} and the existence of the
structure was explained by the net energy gain per encapsulated
fullerenes
\cite{ZerbettoJPC,RochefortPRB2003,OkadaPRB,DubayPRB2004}. Peapods
turned out to be a model system to study molecular interactions such
as fullerene polymerization \cite{PichlerPRL2001} and guest-host
interactions \cite{PfeifferPRB2004} and they are also the precursors
to double-wall carbon nanotubes \cite{BandowCPL2001,PfeifferPRL2003}
which enabled the growth of $^{13}$C isotope enriched inner tubes
\cite{SimonPRL2005}. Concerning applications, peapods were found to
show ambipolar field transistor effect \cite{ShinoharaAPL2002} and
modified field emission characteristics as compared to empty SWCNTs
\cite{ObraztsovKB2005}.

An intriguing class of peapods is that with encapsulated magnetic
fullerenes such as metallofullerenes with magnetic rare-earth ions
\cite{ShinoharaAPL2002}, the endohedral N@C$_{60}$
\cite{SimonCPL2004}, and C$_{59}$N \cite{SimonPRL2006}. Peapods with
magnetic fullerenes can be used to study the electronic properties
of the tubes \cite{SimonPRL2006} and may be suitable for quantum
information processing \cite{HarneitPRA,HarneitPSS} and they are
thought to be good candidates for magnetic force microscopy
cantilevers. The endohedral N@C$_{60}$ fullerene is itself a unique
molecule as it contains an atomic nitrogen in the electron spin
S=3/2 configuration \cite{WeidingerPRL}. The atomic nitrogen weakly
interacts with its environment resulting in long electron
spin-lattice relaxation times, $T_1$s \cite{MehringKB2000}. However,
the N@C$_{60}$ molecule is very sensitive to temperature and
annealing to $\sim$ 500 K was reported to irreversibly destroy it
through the escape of the nitrogen \cite{WaiblingerPRB}. This fact
motivated us to study the temperature stability of N@C$_{60}$ when
it is encapsulated inside SWCNTs. The modification to the electronic
state of the molecule upon encapsulation is also of interest as it
can yield information about the fullerene-tube interaction and about
the electronic structure of the tubes themselves.

Here, we report on high temperature electron-spin resonance (ESR)
spectroscopy on the encapsulated N@C$_{60}$. We find that the
molecule decays much slower than its pristine counterpart in the
crystalline form, which indicates that the presence of the nanotubes
stabilize the molecule. $T_1$ of N@C$_{60}$ in the peapod form is
significantly shorter than in the crystalline form due to the
interaction with the conduction electrons on the nanotubes.

\section{\label{sec:level1}Experimental\protect}

We prepared peapods from commercial purified SWCNTs (Nanocarblab,
Moscow, Russia, purity 50 wt\%) and N@C$_{60}$:C$_{60}$ fullerenes
with an N@C$_{60}$ concentration of 400 ppm. The endohedral
fullerene was produced by the N implantation method
\cite{WeidingerPRL} followed by high performance liquid
chromatography, which purifies and concentrates the material. A
suspension of SWCNT in toluene containing dissolved
N@C$_{60}$:C$_{60}$ was sonicated for 2 hours, filtered with a 0.4
micron pore size filter and dried at room temperature to obtain
samples in the form of bucky-papers. Raman, electron-energy loss
spectroscopies, and X-ray diffraction studies indicate that sizable
filling can be obtained with this route \cite{SimonCPL2004}. After
degassing at room temperature, the bucky-paper samples were ground
to enable penetration of the exciting microwaves and to avoid
microwave losses. In the following, we refer to the starting
N@C$_{60}$:C$_{60}$ as "crystalline" and to the encapsulated
material as "peapod". Both kinds of samples were sealed in quartz
tubes under He atmosphere. Experiments were carried out using a
Bruker Elexsys E500 spectrometer with a TE011 microwave cavity
equipped with gas-flow inserts for the low (5-300 K) and high
(300-620 K) temperature measurements. The ESR signal intensity,
which is proportional to the number of spins observed, was
determined by fitting Lorentzian curves to the data.

\section{Results and Discussion}

\begin{figure}[tbp]
\includegraphics[width=0.9\hsize]{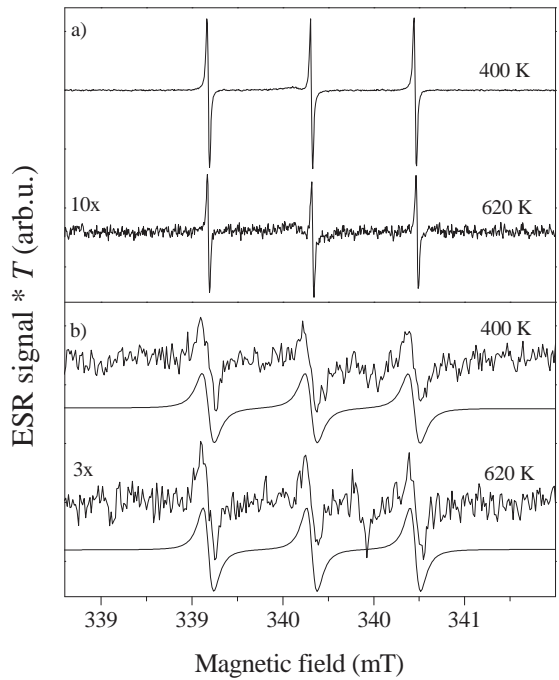}
\caption{ESR spectra of the crystalline (a) and peapod (b)
N@C$_{60}$:C$_{60}$ multiplied by the temperature. The data are
shown on different scales for the two kinds of materials. Note the
magnified vertical scale for the high temperature data. Solid curves
are fit to the data.} \label{spectra}
\end{figure}

The ESR spectra of the crystalline and peapod N@C$_{60}$:C$_{60}$
materials are shown for different temperatures in Fig.
\ref{spectra}. A triplet signal, which is characteristic for the
hyperfine interaction of the N spins with the $^{14}$N nucleus with
nuclear spin $I=1$ is observed for both kinds of samples. The three
unpaired electrons on the 2p$^{3}$ nitrogen atomic orbitals are in a
high spin, S=3/2 state configuration. The zero-field splitting of
the electron Zeeman levels is small due to the high symmetry of the
fullerene cage and it can be observed in spin-echo ESR experiments
only \cite{DinseJPCB1999}. The isotropic nuclear hyperfine coupling,
$A_{\text{iso}}=0.565$ mT of N@C$_{60}$ is uniquely large due to the
compression of the nitrogen orbitals, which unambiguously identifies
the observation of this molecule \cite{WeidingerPRL}.

Besides the triplet, the peapod spectra contains a broad background
(not shown) due to the inevitable presence of ferromagnetic Ni:Y
catalysts in the SWCNT samples. However, the peapod spectrum does
not contain any impurity line around $g=2$ as observed in the early
studies \cite{SimonCPL2004}, which attests the high SWCNT sample
purity. The ESR signal intensity observed for the peapod sample was
compared to that in the crystalline sample, which enables to
determine the amount of N@C$_{60}$ in the peapods. We found that
fullerene content in the peapod is $\sim$ 2--3 weight-percent, in
agreement with the previous studies
\cite{SimonCPL2004,SimonPRL2005}.

\begin{figure}[tbp]
\includegraphics[width=0.9\hsize]{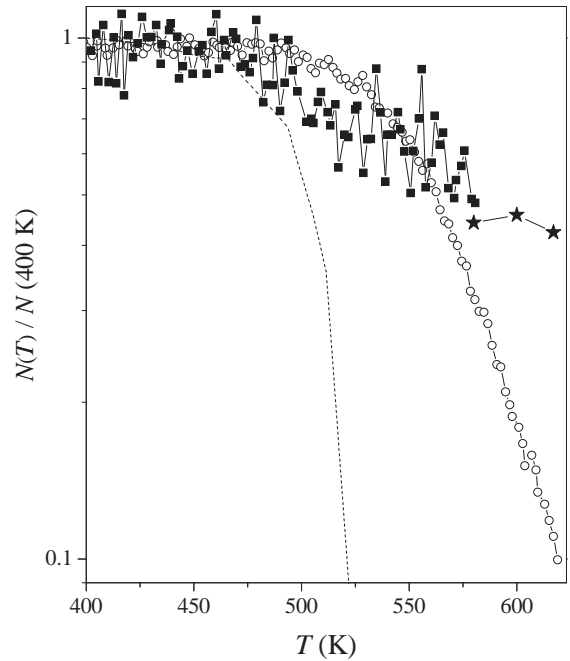}
\caption{Temperature dependence of the number of N@C$_{60}$ spins,
$N(T)$ normalized to the 400 K values, $\bigcirc$: crystalline,
$\blacksquare$: peapod. The longer averaged data for the peapod
material at 580, 600, and 620 K are shown with asterisks. Dashed
curve shows similar measurements on the crystalline material from
Ref. \cite{WaiblingerPRB}. Note the logarithmic vertical scale.}
\label{intensitylog}
\end{figure}

The number of N@C$_{60}$ spins, $N(T)$, is proportional to the
product of the ESR signal intensity and the temperature as the
magnetic susceptibility of the N@C$_{60}$ spins follows a Curie
temperature dependence. Since we are interested in the change of
$N(T)$ as a function of temperature, normalizing it to a well
defined temperature value allows a precise monitoring of the decay
of N@C$_{60}$. In Fig. \ref{intensitylog}., we show the temperature
dependence of the number of N@C$_{60}$ spins in both kinds of
samples normalized to the value at 400 K. For comparison, we show
the corresponding data from Ref. \cite{WaiblingerPRB}. The annealing
speed was 90 s/2 K, that is identical to the heating protocol used
in Ref. \cite{WaiblingerPRB}. At 580, 600, and 620 K, we performed
longer data acquisition thus less temperature points were taken to
maintain the annealing protocol. We observe a clear drop in the
number of N@C$_{60}$ for the crystalline material when heated above
$\sim$ 550 K. However, we observe that the decay is less sharp and
that it occurs at about 50 K higher temperature than that observed
previously \cite{WaiblingerPRB}. We have no clear explanation for
this difference between ours and the previous studies. We carefully
checked our thermometry and we used a large flux of exchange gas. In
addition, ex-situ annealing, i.e. heating both samples outside the
ESR cavity in a furnace while approximately following the above
annealing protocol, gave identical decay curves as the in-situ
result.

The important observation is that the number of N@C$_{60}$ does not
decay in the peapod sample as fast as in the crystalline material.
We observe about relatively three times as much N@C$_{60}$ in the
peapod material at 620 K than in the crystalline material. This
effect is also apparent in Fig. \ref{spectra}. where the
corresponding spectra are shown. Although the noise in the peapod
data limits the conclusions, the decay of N@C$_{60}$ appears to
start already at 500 K but there is no sharp decay such as that
observed for the crystalline material. The amount of N@C$_{60}$
decreases rather smoothly with increasing temperature, however we
could not follow this above 620 K due to technical limitations.

In the following, we discuss the origin of the enhanced thermal
stability of N@C$_{60}$ inside nanotubes. It was proposed in Ref.
\cite{WaiblingerPRB} that the atomic nitrogen escapes from the
fullerene cage by forming bonds with two neighboring carbon atoms
from the inside and by swinging through the bonds to the outside of
the fullerene. This was supported by the observation of enhanced
stability of the encaged nitrogen when the fullerene was
functionalized, which effectively suppresses the probability of this
escape path. \textit{Ab-initio} electronic structure calculations on
the peapods indicate a hybridization of the orbitals on the
fullerenes and the nanotubes \cite{DubayPRB2004}. Raman spectroscopy
on the peapods provided experimental evidence for hybridization and
a partial charge transfer between the nanotubes and the fullerenes
\cite{PichlerPRL2001}. We suggest that these effects suppress the
inside-the-cage bond formation for the peapod N@C$_{60}$ similar to
functionalization of the molecules.

In addition to the modified electronic structure of fullerenes, the
peapod geometry may also play a role in the enhanced thermal
stability of N@C$_{60}$. The one-dimensional lattice constant of
fullerenes inside the tubes is 0.97 nm \cite{HiraharaPRB} that is
about half-way between the 1.002 nm fullerene-fullerene spacing in
crystalline C$_{60}$ \cite{HeineyPRL1991} and 0.961 nm for
polymerized \textit{fcc} C$_{60}$ \cite{IwasaSCI1994}. The eight
voids between the fullerenes and the host nanotube wall are
significantly smaller than the so-called octahedral and tetrahedral
voids in crystalline C$_{60}$. This, combined with the hindered
rotation of the fullerenes \cite{SimonPRL2006}, probably limits the
above described escape process of nitrogen simply by geometrically
limiting the available room for the above described swinging-out
process. For both mechanisms, the Gaussian distribution of tube
diameters with an average of 1.4 nm and 0.1 nm variance explains why
the thermally induced decay of N@C$_{60}$ is spread out in
temperature: we expect that both the electronic modification of the
fullerenes through the fullerene-nanotube interaction and also the
geometrical effect are strongly influenced by the diameter of the
host tubes.

\begin{figure}[tbp]
\includegraphics[width=0.9\hsize]{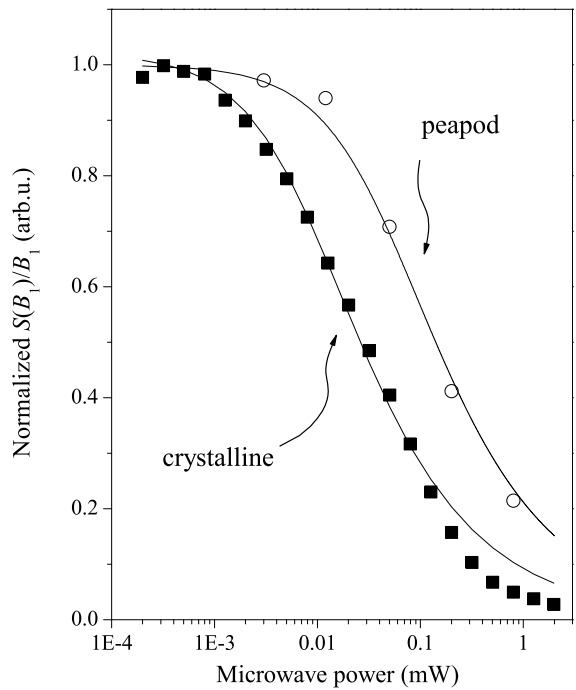}
\caption{Saturation curves of N@C$_{60}$ and peapod N@C$_{60}$ at
300 K. The ESR intensities are divided by the microwave magnetic
field, $B_1$ and are normalized to the values at the lowest power.
Solid curves are calculated saturation curves with Eq.
\ref{Theoretical_Saturation}. and the parameters given in Table
\ref{Relaxation_Table}. Note that the solid curve for the
crystalline material is calculated with relaxation times determined
in spin-echo measurements in Ref. \cite{HarneitPRA} with no further
adjustable parameters.} \label{saturation}
\end{figure}

In general, the spin-lattice relaxation time of a paramagnetic spin
can be used to study the electronic structure of its environment.
Well known example of this is the Korringa-relaxation, i.e.
$1/T_1T=$const. of nuclear spins when these are embedded in metals
\cite{SlichterBook}. In our case, the $T_1$ of encapsulated
N@C$_{60}$ can be studied using saturated ESR measurement
\cite{PortisPR1953}. In this method, the ESR signal intensity is
progressively saturated upon increasing microwave power and the ESR
signal intensity follows \cite{PortisPR1953,SlichterBook}:

\begin{eqnarray}
S(p) \propto \frac{\sqrt{p}}{\sqrt{1+C Q \gamma^2 p T_{1} T_2}}
\label{Theoretical_Saturation}
\end{eqnarray}

\noindent where $\gamma_{\text{e}}/2 \pi=28.0 \text{ GHz/T}$ is the
electron gyromagnetic ratio, $p$ is the microwave power, $T_2$ is
the spin-spin relaxation time of the individual spin-packets, $C$ is
a constant which depends on the microwave cavity mode with quality
factor Q and describes how large microwave magnetic field,
$B_1=\sqrt{C Q p}$, is produced for a power of $p$. Taking into
account that only one of the two circularly polarized magnetic field
components of the linearly polarized field is ESR active we obtain
for the TE011 cavity $C= 2.2 \cdot 10^{-12} \text{T}^2/\text{W}$
\cite{PooleBook}. In Fig. \ref{saturation}. we show the saturated
ESR results for the crystalline and the peapod material at 300 K.

\begin{table}
\begin{center}
\begin{tabular}{lllll}
\hline \hline
& $T_1$ [$\mu$s] & &$T_2$  [$\mu$s] &\\
\hline crystalline \cite{HarneitPRA} &120  &  & 5 & \\
peapod  & 13 & & 13 & \\
\hline \hline
\end{tabular}
\caption{Room temperature spin-lattice and spin-spin relaxation
times for the crystalline and peapod materials used to calculate the
saturated ESR data. We used the experimental quality factors of
$Q=3000$ and 2000 for the crystalline and peapod materials,
respectively. The $T_2=\text{5 }\mu \text{s}$ of the crystalline
material was obtained for our 400 ppm sample from the $T_2=\text{20
}\mu \text{s}$ for a 100 ppm N@C$_{60}$:C$_{60}$ sample.}
\label{Relaxation_Table}
\end{center}
\end{table}

Clearly, the ESR signal of the peapod material saturates at larger
microwave powers, which indicates a shorter $T_1$ relaxation time.
To obtain $T_1$ values from the saturation curves using Eq.
\ref{Theoretical_Saturation}, the value for $T_2$ has to be known.
$T_2$ is the spin-spin relaxation time of individual spin-packets
that is given by the dipolar interaction of like-spins provided
$T_1$ is long enough and does not give a homogeneous broadening
\cite{AbragamBook}. If this is the case, $T_2$ can be determined
from the dipolar interaction strength of the like-spins and is
inversely proportional to their concentration. $T_2=\text{5 }\mu
\text{s}$ is obtained for the 400 ppm N@C$_{60}$:C$_{60}$
crystalline material from the $T_2=\text{20 }\mu \text{s}$ for a 100
ppm N@C$_{60}$:C$_{60}$ sample \cite{HarneitPRA}. As shown in Table
\ref{Relaxation_Table}., $T_1$ is much longer for the crystalline
material so no homogeneous broadening of the spin-packets occurs.
This situation is reversed for the peapod material: the low
concentration of the like nitrogen spins would give a long $T_2 \sim
\text{250 }\mu \text{s}$ based on the $\sim 2$ \% fullerene weight
percentage in the peapod. However, $T_1$ is shorter than that value,
which gives a homogeneous broadening of the individual spin-packets
and sets $T_1=T_2$. As we show below, this holds down to the lowest
temperatures. It is interesting to note here, that a similar
situation was encountered for another diluted magnetic fullerene
peapod system, the C$_{59}$N:C$_{60}$, where the rapid $T_1$
relaxation causes a homogeneous broadening \cite{SimonPRL2006}.

We show the simulated saturation curves in Fig. \ref{saturation}.
with the parameters given in Table \ref{Relaxation_Table}. The
excellent agreement for the measured and calculated saturation
curves for the crystalline material shows that the saturated ESR
measurement can be used to determine values for $T_1$ and $T_2$,
although it lacks the direct access to these values such as the
spin-echo ESR method. It thus justifies the use of the saturated ESR
method to determine the $T_1$ and $T_2$ values for the peapod system
as well.

\begin{figure}[tbp]
\includegraphics[width=0.9\hsize]{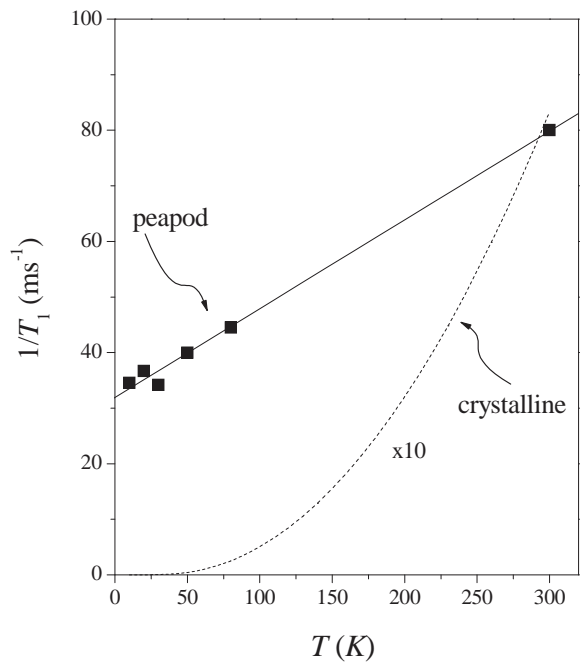}
\caption{Temperature dependence of the spin-lattice relaxation rate,
$1/T_1$, for peapod N@C$_{60}$. We show the corresponding data for
the crystalline material from Ref. \cite{MehringKB2000} magnified by
10 with a dashed curve. Solid line is a guide to the eye.}
\label{Tdep_T1}
\end{figure}

We determined the temperature dependence of $T_1$ from the saturated
ESR measurements in the 10-80 K temperature range in addition to the
300 K data. Measurements at other temperatures were hindered by
spectrometer stability. We show the $1/T_1$ data in Fig.
\ref{Tdep_T1}. for the peapod material. We also show the same data
for the crystalline material from Ref. \cite{MehringKB2000}. The
$1/T_1$ spin-lattice relaxation rate is roughly one order of
magnitude larger for the peapod material than for the crystalline at
room temperature. It is even larger at 10 K for the peapod than the
corresponding value for the crystalline material. The temperature
dependence of $1/T_1$ is also different for the two kinds of
compounds: for the peapod it decreases with temperature to a
residual value, whereas it vanishes exponentially for the
crystalline material \cite{MehringKB2000}. It was proposed in Ref.
\cite{MehringKB2000} that $T_1$ in crystalline N@C$_{60}$ is given
by the quantum oscillator motion of the encaged nitrogen through the
modulation of the hyperfine field, which results in the exponential
freezing-out of the relaxation rate.

The enhanced relaxation rate in the peapod material is given by
additional relaxation mechanisms. We expect that the hyperfine
relaxation is not significantly different in the peapod material as
the isotropic hyperfine coupling is unchanged. In addition, the
altered rotational dynamics of the molecule inside the tubes cannot
have a significant effect on the hyperfine relaxation, contrary to
that was concluded previously \cite{CorziliusKB2005}. Possible
mechanisms for the relaxation are coupling to conduction electrons
on the tubes and paramagnetic relaxation from defects on the tubes
and from the transition metal catalyst particles, which are
inevitably present in the nanotube samples. The latter relaxation
mechanism could explain the presence of the residual relaxation rate
as it is inversely proportional to the temperature
\cite{AbragamBook}. This contribution is expected to be negligible
at room temperature, therefore the factor 10 enhancement of the room
temperature relaxation rate for the peapod is suggested to originate
from the coupling of the nitrogen spins to the conduction electrons
on the tubes, i.e. from a Korringa relaxation. It is important to
note here, that we do not observe a multi-component saturation, i.e.
there seems to be a single $T_1$ time for the encapsulated
N@C$_{60}$.

An approximate value for the Korringa-relaxation related relaxation
rate is obtained from the 90 \% of the room temperature $1/T_1$
value, that gives $1/T_1(\text{300 K})=0.24 \text{
ms}^{-1}\text{K}^{-1} $. The theory of Korringa relaxation can be
used to obtain the $J=0.9 $ meV coupling constant between the
localized spins and the conduction electrons \cite{SlichterBook}:

\begin{center}
\begin{eqnarray}
\frac{1}{T_{1}T}=\left( \frac{4\pi k_{B}}{\hbar }\right)
J^{2}\overline{n}(E_{F})^{2} \label{Korringa_relation}
\end{eqnarray}
\end{center}

\noindent where $\overline{n}%
(E_{F})=0.014 \text{ states/eV/atom}$ is the DOS at the Fermi level for a $%
d\approx 1.4$ nm metallic tube in the tight-binding approximation \cite%
{DresselhausTubesNew}. The current value of $J$ is significantly
lower than the $J=11 $ meV found for the coupling of C$_{59}$N spins
and the itinerant electrons on the nanotubes. This difference
originates from the encaged nature of the nitrogen spins for
N@C$_{60}$. Nevertheless, the above mentioned uniformity of the
saturation curves suggests that only one kind of nanotubes, namely
metallic ones are detected using magnetic resonance, which confirms
the earlier observations using NMR \cite{SingerPRL2005} and ESR
\cite{SimonPRL2006}. The fact that the spin-lattice relaxation rate
of N@C$_{60}$ remains finite at the lowest temperature puts a severe
limit on the applicability of the NC$_{60}$ peapod system for
quantum information processing.

\section{Conclusions}
In summary, we found that the N@C$_{60}$ molecule is more stable in
peapod than in crystalline form. This was suggested to result from
the modified electronic structure of fullerenes inside the tubes or
alternatively from the compact packing of the fullerenes inside the
nanotubes. The spin-lattice relaxation time of the N@C$_{60}$ spins
is much shorter in the peapod form due to the interaction with the
electrons on the nanotubes. This indicates a finite density of
states at the Fermi level of most of the nanotubes.

\section{Acknowledgement}
The authors are grateful to A. J\'{a}nossy and A. Rockenbauer for
stimulating discussions. We thank K.-P. Dinse for providing the
endohedral fullerenes. We acknowledge support by the Hungarian State
Grants (OTKA) No. TS049881, F61733, NK60984, and T046953. FS
acknowledges the Zolt\'{a}n Magyary and the Bolyai fellowships for
support. The work in Lausanne was supported by the Swiss National
Science Foundation and by the European research network IMPRESS.

$^{\ast }$ Corresponding author: ferenc.simon@univie.ac.at


\begin{thebibliography}{33}
\expandafter\ifx\csname
natexlab\endcsname\relax\def\natexlab#1{#1}\fi
\expandafter\ifx\csname bibnamefont\endcsname\relax
  \def\bibnamefont#1{#1}\fi
\expandafter\ifx\csname bibfnamefont\endcsname\relax
  \def\bibfnamefont#1{#1}\fi
\expandafter\ifx\csname citenamefont\endcsname\relax
  \def\citenamefont#1{#1}\fi
\expandafter\ifx\csname url\endcsname\relax
  \def\url#1{\texttt{#1}}\fi
\expandafter\ifx\csname urlprefix\endcsname\relax\def\urlprefix{URL
}\fi \providecommand{\bibinfo}[2]{#2}
\providecommand{\eprint}[2][]{\url{#2}}

\bibitem[{\citenamefont{Iijima and Ichihashi}(1993)}]{IijimaNat1993}
\bibinfo{author}{\bibfnamefont{S.}~\bibnamefont{Iijima}} \bibnamefont{and}
  \bibinfo{author}{\bibfnamefont{T.}~\bibnamefont{Ichihashi}},
  \bibinfo{journal}{Nature} \textbf{\bibinfo{volume}{363}},
  \bibinfo{pages}{603} (\bibinfo{year}{1993}).

\bibitem[{\citenamefont{Bethune et~al.}(1993)\citenamefont{Bethune, Kiang,
  DeVries, Gorman, R., and Beyers}}]{BethuneNat1993}
\bibinfo{author}{\bibfnamefont{D.~S.} \bibnamefont{Bethune}},
  \bibinfo{author}{\bibfnamefont{C.~H.} \bibnamefont{Kiang}},
  \bibinfo{author}{\bibfnamefont{M.~S.} \bibnamefont{DeVries}},
  \bibinfo{author}{\bibfnamefont{G.}~\bibnamefont{Gorman}},
  \bibinfo{author}{\bibfnamefont{S.}~\bibnamefont{R.}}, \bibnamefont{and}
  \bibinfo{author}{\bibfnamefont{R.}~\bibnamefont{Beyers}},
  \bibinfo{journal}{Nature} \textbf{\bibinfo{volume}{363}},
  \bibinfo{pages}{605} (\bibinfo{year}{1993}).

\bibitem[{\citenamefont{Smith et~al.}(1998)\citenamefont{Smith, Monthioux, and
  Luzzi}}]{SmithNat}
\bibinfo{author}{\bibfnamefont{B.~W.} \bibnamefont{Smith}},
  \bibinfo{author}{\bibfnamefont{M.}~\bibnamefont{Monthioux}},
  \bibnamefont{and} \bibinfo{author}{\bibfnamefont{D.~E.} \bibnamefont{Luzzi}},
  \bibinfo{journal}{Nature} \textbf{\bibinfo{volume}{396}},
  \bibinfo{pages}{323} (\bibinfo{year}{1998}).

\bibitem[{\citenamefont{Berber et~al.}(2002)\citenamefont{Berber, Kwon, and
  Tom\'{a}nek}}]{TomanekPRL}
\bibinfo{author}{\bibfnamefont{S.}~\bibnamefont{Berber}},
  \bibinfo{author}{\bibfnamefont{Y.-K.} \bibnamefont{Kwon}}, \bibnamefont{and}
  \bibinfo{author}{\bibfnamefont{D.}~\bibnamefont{Tom\'{a}nek}},
  \bibinfo{journal}{Phys. Rev. Lett.} \textbf{\bibinfo{volume}{88}},
  \bibinfo{pages}{185502} (\bibinfo{year}{2002}).

\bibitem[{\citenamefont{Melle-Franco et~al.}(2003)\citenamefont{Melle-Franco,
  Kuzmany, and Zerbetto}}]{ZerbettoJPC}
\bibinfo{author}{\bibfnamefont{M.}~\bibnamefont{Melle-Franco}},
  \bibinfo{author}{\bibfnamefont{H.}~\bibnamefont{Kuzmany}}, \bibnamefont{and}
  \bibinfo{author}{\bibfnamefont{F.}~\bibnamefont{Zerbetto}},
  \bibinfo{journal}{J. Phys. Chem. B} \textbf{\bibinfo{volume}{109}},
  \bibinfo{pages}{6986} (\bibinfo{year}{2003}).

\bibitem[{\citenamefont{Rochefort}(2003)}]{RochefortPRB2003}
\bibinfo{author}{\bibfnamefont{A.}~\bibnamefont{Rochefort}},
  \bibinfo{journal}{Appl. Magn. Reson.} \textbf{\bibinfo{volume}{67}},
  \bibinfo{pages}{115401–1–7} (\bibinfo{year}{2003}).

\bibitem[{\citenamefont{Otani et~al.}(2003)\citenamefont{Otani, Okada, and
  Oshiyama}}]{OkadaPRB}
\bibinfo{author}{\bibfnamefont{M.}~\bibnamefont{Otani}},
  \bibinfo{author}{\bibfnamefont{S.}~\bibnamefont{Okada}}, \bibnamefont{and}
  \bibinfo{author}{\bibfnamefont{A.}~\bibnamefont{Oshiyama}},
  \bibinfo{journal}{Phys. Rev. B} \textbf{\bibinfo{volume}{68}},
  \bibinfo{pages}{125424} (\bibinfo{year}{2003}).

\bibitem[{\citenamefont{Dubay and Kresse}(2004)}]{DubayPRB2004}
\bibinfo{author}{\bibfnamefont{O.}~\bibnamefont{Dubay}} \bibnamefont{and}
  \bibinfo{author}{\bibfnamefont{G.}~\bibnamefont{Kresse}},
  \bibinfo{journal}{Phys. Rev. B} \textbf{\bibinfo{volume}{70}},
  \bibinfo{pages}{165424} (\bibinfo{year}{2004}).

\bibitem[{\citenamefont{Pichler et~al.}(2001)\citenamefont{Pichler, Kuzmany,
  Kataura, and Achiba}}]{PichlerPRL2001}
\bibinfo{author}{\bibfnamefont{T.}~\bibnamefont{Pichler}},
  \bibinfo{author}{\bibfnamefont{H.}~\bibnamefont{Kuzmany}},
  \bibinfo{author}{\bibfnamefont{H.}~\bibnamefont{Kataura}}, \bibnamefont{and}
  \bibinfo{author}{\bibfnamefont{Y.}~\bibnamefont{Achiba}},
  \bibinfo{journal}{Phys. Rev. Lett.} \textbf{\bibinfo{volume}{87}},
  \bibinfo{pages}{267401} (\bibinfo{year}{2001}).

\bibitem[{\citenamefont{Pfeiffer et~al.}(2004)\citenamefont{Pfeiffer, Kuzmany,
  Pichler, Kataura, Achiba, Melle-Franco, and Zerbetto}}]{PfeifferPRB2004}
\bibinfo{author}{\bibfnamefont{R.}~\bibnamefont{Pfeiffer}},
  \bibinfo{author}{\bibfnamefont{H.}~\bibnamefont{Kuzmany}},
  \bibinfo{author}{\bibfnamefont{T.}~\bibnamefont{Pichler}},
  \bibinfo{author}{\bibfnamefont{H.}~\bibnamefont{Kataura}},
  \bibinfo{author}{\bibfnamefont{Y.}~\bibnamefont{Achiba}},
  \bibinfo{author}{\bibfnamefont{M.}~\bibnamefont{Melle-Franco}},
  \bibnamefont{and} \bibinfo{author}{\bibfnamefont{F.}~\bibnamefont{Zerbetto}},
  \bibinfo{journal}{Phys. Rev. B} \textbf{\bibinfo{volume}{69}},
  \bibinfo{pages}{035404} (\bibinfo{year}{2004}).

\bibitem[{\citenamefont{Bandow et~al.}(2001)\citenamefont{Bandow, Takizawa,
  Hirahara, Yudasaka, and Iijima}}]{BandowCPL2001}
\bibinfo{author}{\bibfnamefont{S.}~\bibnamefont{Bandow}},
  \bibinfo{author}{\bibfnamefont{M.}~\bibnamefont{Takizawa}},
  \bibinfo{author}{\bibfnamefont{K.}~\bibnamefont{Hirahara}},
  \bibinfo{author}{\bibfnamefont{M.}~\bibnamefont{Yudasaka}}, \bibnamefont{and}
  \bibinfo{author}{\bibfnamefont{S.}~\bibnamefont{Iijima}},
  \bibinfo{journal}{Chem. Phys. Lett.} \textbf{\bibinfo{volume}{337}},
  \bibinfo{pages}{48} (\bibinfo{year}{2001}).

\bibitem[{\citenamefont{Pfeiffer et~al.}(2003)\citenamefont{Pfeiffer, Kuzmany,
  Kramberger, Schaman, Pichler, Kataura, Achiba, K{\"u}rti, and
  Z{\'o}lyomi}}]{PfeifferPRL2003}
\bibinfo{author}{\bibfnamefont{R.}~\bibnamefont{Pfeiffer}},
  \bibinfo{author}{\bibfnamefont{H.}~\bibnamefont{Kuzmany}},
  \bibinfo{author}{\bibfnamefont{C.}~\bibnamefont{Kramberger}},
  \bibinfo{author}{\bibfnamefont{C.}~\bibnamefont{Schaman}},
  \bibinfo{author}{\bibfnamefont{T.}~\bibnamefont{Pichler}},
  \bibinfo{author}{\bibfnamefont{H.}~\bibnamefont{Kataura}},
  \bibinfo{author}{\bibfnamefont{Y.}~\bibnamefont{Achiba}},
  \bibinfo{author}{\bibfnamefont{J.}~\bibnamefont{K{\"u}rti}},
  \bibnamefont{and}
  \bibinfo{author}{\bibfnamefont{V.}~\bibnamefont{Z{\'o}lyomi}},
  \bibinfo{journal}{Phys. Rev. Lett.} \textbf{\bibinfo{volume}{90}},
  \bibinfo{pages}{225501} (\bibinfo{year}{2003}).

\bibitem[{\citenamefont{Simon et~al.}(2005)\citenamefont{Simon, Kramberger,
  Pfeiffer, Kuzmany, Z\'{o}lyomi, K\"{u}rti, Singer, and
  Alloul}}]{SimonPRL2005}
\bibinfo{author}{\bibfnamefont{F.}~\bibnamefont{Simon}},
  \bibinfo{author}{\bibfnamefont{C.}~\bibnamefont{Kramberger}},
  \bibinfo{author}{\bibfnamefont{R.}~\bibnamefont{Pfeiffer}},
  \bibinfo{author}{\bibfnamefont{H.}~\bibnamefont{Kuzmany}},
  \bibinfo{author}{\bibfnamefont{V.}~\bibnamefont{Z\'{o}lyomi}},
  \bibinfo{author}{\bibfnamefont{J.}~\bibnamefont{K\"{u}rti}},
  \bibinfo{author}{\bibfnamefont{P.~M.} \bibnamefont{Singer}},
  \bibnamefont{and} \bibinfo{author}{\bibfnamefont{H.}~\bibnamefont{Alloul}},
  \bibinfo{journal}{Phys. Rev. Lett.} \textbf{\bibinfo{volume}{95}},
  \bibinfo{pages}{017401} (\bibinfo{year}{2005}).

\bibitem[{\citenamefont{Shimada et~al.}(2002)\citenamefont{Shimada, Okazaki,
  Taniguchi, Sugai, Shinohara, Suenaga, Ohno, Mizuno, Kishimoto, and
  Mizutani}}]{ShinoharaAPL2002}
\bibinfo{author}{\bibfnamefont{T.}~\bibnamefont{Shimada}},
  \bibinfo{author}{\bibfnamefont{T.}~\bibnamefont{Okazaki}},
  \bibinfo{author}{\bibfnamefont{R.}~\bibnamefont{Taniguchi}},
  \bibinfo{author}{\bibfnamefont{T.}~\bibnamefont{Sugai}},
  \bibinfo{author}{\bibfnamefont{H.}~\bibnamefont{Shinohara}},
  \bibinfo{author}{\bibfnamefont{K.}~\bibnamefont{Suenaga}},
  \bibinfo{author}{\bibfnamefont{Y.}~\bibnamefont{Ohno}},
  \bibinfo{author}{\bibfnamefont{S.}~\bibnamefont{Mizuno}},
  \bibinfo{author}{\bibfnamefont{S.}~\bibnamefont{Kishimoto}},
  \bibnamefont{and} \bibinfo{author}{\bibfnamefont{T.}~\bibnamefont{Mizutani}},
  \bibinfo{journal}{Appl. Phys. Lett.} \textbf{\bibinfo{volume}{81}},
  \bibinfo{pages}{4067} (\bibinfo{year}{2002}).

\bibitem[{\citenamefont{Lyashenko et~al.}(2005)\citenamefont{Lyashenko,
  Obraztsov, Simon, Kuzmany, Obraztsova, Svirko, and
  Jefimovs}}]{ObraztsovKB2005}
\bibinfo{author}{\bibfnamefont{D.~A.} \bibnamefont{Lyashenko}},
  \bibinfo{author}{\bibfnamefont{A.~N.} \bibnamefont{Obraztsov}},
  \bibinfo{author}{\bibfnamefont{F.}~\bibnamefont{Simon}},
  \bibinfo{author}{\bibfnamefont{H.}~\bibnamefont{Kuzmany}},
  \bibinfo{author}{\bibfnamefont{E.~D.} \bibnamefont{Obraztsova}},
  \bibinfo{author}{\bibfnamefont{Y.~P.} \bibnamefont{Svirko}},
  \bibnamefont{and} \bibinfo{author}{\bibfnamefont{K.}~\bibnamefont{Jefimovs}},
  \bibinfo{journal}{AIP Conf. Proc.} \textbf{\bibinfo{volume}{786}},
  \bibinfo{pages}{301} (\bibinfo{year}{2005}).

\bibitem[{\citenamefont{Simon et~al.}(2004)\citenamefont{Simon, Kuzmany, Rauf,
  Pichler, Bernardi, Peterlik, Korecz, F\"ul\"op, and
  J\'anossy}}]{SimonCPL2004}
\bibinfo{author}{\bibfnamefont{F.}~\bibnamefont{Simon}},
  \bibinfo{author}{\bibfnamefont{H.}~\bibnamefont{Kuzmany}},
  \bibinfo{author}{\bibfnamefont{H.}~\bibnamefont{Rauf}},
  \bibinfo{author}{\bibfnamefont{T.}~\bibnamefont{Pichler}},
  \bibinfo{author}{\bibfnamefont{J.}~\bibnamefont{Bernardi}},
  \bibinfo{author}{\bibfnamefont{H.}~\bibnamefont{Peterlik}},
  \bibinfo{author}{\bibfnamefont{L.}~\bibnamefont{Korecz}},
  \bibinfo{author}{\bibfnamefont{F.}~\bibnamefont{F\"ul\"op}},
  \bibnamefont{and}
  \bibinfo{author}{\bibfnamefont{A.}~\bibnamefont{J\'anossy}},
  \bibinfo{journal}{Chem. Phys. Lett.} \textbf{\bibinfo{volume}{383}},
  \bibinfo{pages}{362} (\bibinfo{year}{2004}).

\bibitem[{\citenamefont{Simon et~al.}(2006)\citenamefont{Simon, Kuzmany,
  N\'{a}fr\'{a}di, Feh\'{e}r, Forr\'{o}, F\"{u}l\"{o}p, J\'{a}nossy,
  Rockenbauer, Korecz, Hauke et~al.}}]{SimonPRL2006}
\bibinfo{author}{\bibfnamefont{F.}~\bibnamefont{Simon}},
  \bibinfo{author}{\bibfnamefont{H.}~\bibnamefont{Kuzmany}},
  \bibinfo{author}{\bibfnamefont{B.}~\bibnamefont{N\'{a}fr\'{a}di}},
  \bibinfo{author}{\bibfnamefont{T.}~\bibnamefont{Feh\'{e}r}},
  \bibinfo{author}{\bibfnamefont{L.}~\bibnamefont{Forr\'{o}}},
  \bibinfo{author}{\bibfnamefont{F.}~\bibnamefont{F\"{u}l\"{o}p}},
  \bibinfo{author}{\bibfnamefont{A.}~\bibnamefont{J\'{a}nossy}},
  \bibinfo{author}{\bibfnamefont{A.}~\bibnamefont{Rockenbauer}},
  \bibinfo{author}{\bibfnamefont{L.}~\bibnamefont{Korecz}},
  \bibinfo{author}{\bibfnamefont{F.}~\bibnamefont{Hauke}},
  \bibnamefont{et~al.}, \bibinfo{journal}{Phys. Rev. Lett.}
  \textbf{\bibinfo{volume}{97}}, \bibinfo{pages}{136801}
  (\bibinfo{year}{2006}).

\bibitem[{\citenamefont{Harneit}(2002)}]{HarneitPRA}
\bibinfo{author}{\bibfnamefont{W.}~\bibnamefont{Harneit}},
  \bibinfo{journal}{Phys. Rev. A} \textbf{\bibinfo{volume}{65}},
  \bibinfo{pages}{032322} (\bibinfo{year}{2002}).

\bibitem[{\citenamefont{Harneit et~al.}(2002)\citenamefont{Harneit, Meyer,
  Weidinger, Suter, and Twamley}}]{HarneitPSS}
\bibinfo{author}{\bibfnamefont{W.}~\bibnamefont{Harneit}},
  \bibinfo{author}{\bibfnamefont{C.}~\bibnamefont{Meyer}},
  \bibinfo{author}{\bibfnamefont{A.}~\bibnamefont{Weidinger}},
  \bibinfo{author}{\bibfnamefont{D.}~\bibnamefont{Suter}}, \bibnamefont{and}
  \bibinfo{author}{\bibfnamefont{J.}~\bibnamefont{Twamley}},
  \bibinfo{journal}{Phys. St. Solidi B} \textbf{\bibinfo{volume}{233}},
  \bibinfo{pages}{453} (\bibinfo{year}{2002}).

\bibitem[{\citenamefont{Almeida~Murphy
  et~al.}(1996)\citenamefont{Almeida~Murphy, Pawlik, Weidinger, Höhne, Alcala,
  and Spaeth}}]{WeidingerPRL}
\bibinfo{author}{\bibfnamefont{T.}~\bibnamefont{Almeida~Murphy}},
  \bibinfo{author}{\bibfnamefont{T.}~\bibnamefont{Pawlik}},
  \bibinfo{author}{\bibfnamefont{A.}~\bibnamefont{Weidinger}},
  \bibinfo{author}{\bibfnamefont{M.}~\bibnamefont{Höhne}},
  \bibinfo{author}{\bibfnamefont{R.}~\bibnamefont{Alcala}}, \bibnamefont{and}
  \bibinfo{author}{\bibfnamefont{J.-M.} \bibnamefont{Spaeth}},
  \bibinfo{journal}{Phys. Rev. Lett.} \textbf{\bibinfo{volume}{77}},
  \bibinfo{pages}{1075} (\bibinfo{year}{1996}).

\bibitem[{\citenamefont{Knorr et~al.}(2000)\citenamefont{Knorr, Grupp, Mehring,
  Waiblinger, and Weidinger}}]{MehringKB2000}
\bibinfo{author}{\bibfnamefont{S.}~\bibnamefont{Knorr}},
  \bibinfo{author}{\bibfnamefont{A.}~\bibnamefont{Grupp}},
  \bibinfo{author}{\bibfnamefont{M.}~\bibnamefont{Mehring}},
  \bibinfo{author}{\bibfnamefont{M.}~\bibnamefont{Waiblinger}},
  \bibnamefont{and}
  \bibinfo{author}{\bibfnamefont{A.}~\bibnamefont{Weidinger}},
  \bibinfo{journal}{AIP Conf. Proc.} \textbf{\bibinfo{volume}{544}},
  \bibinfo{pages}{191} (\bibinfo{year}{2000}).

\bibitem[{\citenamefont{Waiblinger et~al.}(2001)\citenamefont{Waiblinger, Lips,
  Harneit, Weidinger, Dietel, and Hirsch}}]{WaiblingerPRB}
\bibinfo{author}{\bibfnamefont{M.}~\bibnamefont{Waiblinger}},
  \bibinfo{author}{\bibfnamefont{K.}~\bibnamefont{Lips}},
  \bibinfo{author}{\bibfnamefont{W.}~\bibnamefont{Harneit}},
  \bibinfo{author}{\bibfnamefont{A.}~\bibnamefont{Weidinger}},
  \bibinfo{author}{\bibfnamefont{E.}~\bibnamefont{Dietel}}, \bibnamefont{and}
  \bibinfo{author}{\bibfnamefont{A.}~\bibnamefont{Hirsch}},
  \bibinfo{journal}{Phys. Rev. B} \textbf{\bibinfo{volume}{64}},
  \bibinfo{pages}{159901} (\bibinfo{year}{2001}).

\bibitem[{\citenamefont{Weiden et~al.}(1999)\citenamefont{Weiden, K\"{a}{\ss},
  and Dinse}}]{DinseJPCB1999}
\bibinfo{author}{\bibfnamefont{N.}~\bibnamefont{Weiden}},
  \bibinfo{author}{\bibfnamefont{H.}~\bibnamefont{K\"{a}{\ss}}},
  \bibnamefont{and} \bibinfo{author}{\bibfnamefont{K.~P.} \bibnamefont{Dinse}},
  \bibinfo{journal}{J. Phys. Chem. B} \textbf{\bibinfo{volume}{103}},
  \bibinfo{pages}{9826} (\bibinfo{year}{1999}).

\bibitem[{\citenamefont{Hirahara et~al.}(2001)\citenamefont{Hirahara, Bandow,
  Kato, Okazaki, Shinohara, and Iijima}}]{HiraharaPRB}
\bibinfo{author}{\bibfnamefont{K.}~\bibnamefont{Hirahara}},
  \bibinfo{author}{\bibfnamefont{K.}~\bibnamefont{Bandow},
  \bibfnamefont{S.~Suenaga}},
  \bibinfo{author}{\bibfnamefont{H.}~\bibnamefont{Kato}},
  \bibinfo{author}{\bibfnamefont{T.}~\bibnamefont{Okazaki}},
  \bibinfo{author}{\bibfnamefont{T.}~\bibnamefont{Shinohara}},
  \bibnamefont{and} \bibinfo{author}{\bibfnamefont{S.}~\bibnamefont{Iijima}},
  \bibinfo{journal}{Phys. Rev. B} \textbf{\bibinfo{volume}{64}},
  \bibinfo{pages}{115420} (\bibinfo{year}{2001}).

\bibitem[{\citenamefont{Heiney et~al.}(1991)\citenamefont{Heiney, Fischer,
  McGhie, Romanow, Denenstein, McCauley, and Smith}}]{HeineyPRL1991}
\bibinfo{author}{\bibfnamefont{P.~A.} \bibnamefont{Heiney}},
  \bibinfo{author}{\bibfnamefont{J.~E.} \bibnamefont{Fischer}},
  \bibinfo{author}{\bibfnamefont{A.~R.} \bibnamefont{McGhie}},
  \bibinfo{author}{\bibfnamefont{W.}~\bibnamefont{Romanow}},
  \bibinfo{author}{\bibfnamefont{A.~M.} \bibnamefont{Denenstein}},
  \bibinfo{author}{\bibfnamefont{J.}~\bibnamefont{McCauley}}, \bibnamefont{and}
  \bibinfo{author}{\bibfnamefont{A.~C.~D.} \bibnamefont{Smith}},
  \bibinfo{journal}{Phys. Rev. Lett.} \textbf{\bibinfo{volume}{66}},
  \bibinfo{pages}{2911} (\bibinfo{year}{1991}).

\bibitem[{\citenamefont{Iwasa et~al.}(1994)\citenamefont{Iwasa, Arima, Fleming,
  Siegrist, Zhou, Haddon, Rothberg, Lyons, Carter, Hebard
  et~al.}}]{IwasaSCI1994}
\bibinfo{author}{\bibfnamefont{Y.}~\bibnamefont{Iwasa}},
  \bibinfo{author}{\bibfnamefont{T.}~\bibnamefont{Arima}},
  \bibinfo{author}{\bibfnamefont{R.~M.} \bibnamefont{Fleming}},
  \bibinfo{author}{\bibfnamefont{T.}~\bibnamefont{Siegrist}},
  \bibinfo{author}{\bibfnamefont{O.}~\bibnamefont{Zhou}},
  \bibinfo{author}{\bibfnamefont{R.~C.} \bibnamefont{Haddon}},
  \bibinfo{author}{\bibfnamefont{L.~J.} \bibnamefont{Rothberg}},
  \bibinfo{author}{\bibfnamefont{K.~B.} \bibnamefont{Lyons}},
  \bibinfo{author}{\bibfnamefont{H.~L.} \bibnamefont{Carter}},
  \bibinfo{author}{\bibfnamefont{A.~F.} \bibnamefont{Hebard}},
  \bibnamefont{et~al.}, \bibinfo{journal}{Science}
  \textbf{\bibinfo{volume}{264}}, \bibinfo{pages}{1570} (\bibinfo{year}{1994}).

\bibitem[{\citenamefont{Slichter}(1989)}]{SlichterBook}
\bibinfo{author}{\bibfnamefont{C.~P.} \bibnamefont{Slichter}},
  \emph{\bibinfo{title}{Principles of Magnetic Resonance}}
  (\bibinfo{publisher}{Spinger-Verlag}, \bibinfo{address}{New York},
  \bibinfo{year}{1989}), \bibinfo{edition}{3rd} ed.

\bibitem[{\citenamefont{Portis}(1953)}]{PortisPR1953}
\bibinfo{author}{\bibfnamefont{A.~M.} \bibnamefont{Portis}},
  \bibinfo{journal}{Phys. Rev.} \textbf{\bibinfo{volume}{91}},
  \bibinfo{pages}{1071} (\bibinfo{year}{1953}).

\bibitem[{\citenamefont{Poole}(1983)}]{PooleBook}
\bibinfo{author}{\bibfnamefont{C.~P.} \bibnamefont{Poole}},
  \emph{\bibinfo{title}{Electron Spin Resonance}} (\bibinfo{publisher}{John
  Wiley \& Sons}, \bibinfo{address}{New York}, \bibinfo{year}{1983}),
  \bibinfo{edition}{1983rd} ed.

\bibitem[{\citenamefont{Abragam}(1961)}]{AbragamBook}
\bibinfo{author}{\bibfnamefont{A.}~\bibnamefont{Abragam}},
  \emph{\bibinfo{title}{Principles of Nuclear Magnetism}}
  (\bibinfo{publisher}{Oxford University Press}, \bibinfo{address}{Oxford,
  England}, \bibinfo{year}{1961}).

\bibitem[{\citenamefont{Corzilius et~al.}(2005)\citenamefont{Corzilius, Gembus,
  Dinse, Simon, and Kuzmany}}]{CorziliusKB2005}
\bibinfo{author}{\bibfnamefont{B.}~\bibnamefont{Corzilius}},
  \bibinfo{author}{\bibfnamefont{A.}~\bibnamefont{Gembus}},
  \bibinfo{author}{\bibfnamefont{K.-P.} \bibnamefont{Dinse}},
  \bibinfo{author}{\bibfnamefont{F.}~\bibnamefont{Simon}}, \bibnamefont{and}
  \bibinfo{author}{\bibfnamefont{H.}~\bibnamefont{Kuzmany}},
  \bibinfo{journal}{AIP Conf. Proc.} \textbf{\bibinfo{volume}{786}},
  \bibinfo{pages}{291} (\bibinfo{year}{2005}).

\bibitem[{\citenamefont{Dresselhaus et~al.}(2001)\citenamefont{Dresselhaus,
  Dresselhaus, and Avouris}}]{DresselhausTubesNew}
\bibinfo{author}{\bibfnamefont{M.~S.} \bibnamefont{Dresselhaus}},
  \bibinfo{author}{\bibfnamefont{G.}~\bibnamefont{Dresselhaus}},
  \bibnamefont{and} \bibinfo{author}{\bibfnamefont{P.}~\bibnamefont{Avouris}},
  \emph{\bibinfo{title}{{Carbon Nanotubes: Synthesis, Structure, Properties,
  and Applications}}} (\bibinfo{publisher}{Springer}, \bibinfo{address}{Berlin,
  Heidelberg, New York}, \bibinfo{year}{2001}).

\bibitem[{\citenamefont{Singer et~al.}(2005)\citenamefont{Singer, Wzietek,
  Alloul, Simon, and Kuzmany}}]{SingerPRL2005}
\bibinfo{author}{\bibfnamefont{P.~M.} \bibnamefont{Singer}},
  \bibinfo{author}{\bibfnamefont{P.}~\bibnamefont{Wzietek}},
  \bibinfo{author}{\bibfnamefont{H.}~\bibnamefont{Alloul}},
  \bibinfo{author}{\bibfnamefont{F.}~\bibnamefont{Simon}}, \bibnamefont{and}
  \bibinfo{author}{\bibfnamefont{H.}~\bibnamefont{Kuzmany}},
  \bibinfo{journal}{Phys. Rev. Lett.} \textbf{\bibinfo{volume}{95}},
  \bibinfo{pages}{236403} (\bibinfo{year}{2005}).

\end{thebibliography}

\end{document}